# Observation-dependent suppression and enhancement of two-photon coincidences by tailored losses

Max Ehrhardt, Matthias Heinrich, and Alexander Szameit*

University of Rostock, Institute for Physics, Albert-Einstein-Str. 23, 18059 Rostock, Germany

*Correspondence to: alexander.szameit@uni-rostock.de

**The uncanny ability of multiple particles to interfere with one another is one of the core principles of quantum mechanics, and serves as foundation for quantum information processing. In particular, the interplay of constructive and destructive interference with the characteristic exchange statistics of indistinguishable particles[1–4] give rise to the Hong-Ou-Mandel (HOM) effect[5], where the bunching of bosons can lead to a perfect suppression of two-particle coincidences between the output ports of a balanced beam splitter. Conversely, in the case of two fermions, anti-bunching can systematically enhance these coincidences up to twice the baseline value of distinguishable particles. As such, the respective emergence of dips or peaks in the HOM experiment may at first glance appear to be indicative of the bosonic/fermionic nature of the incident particles. In this work, we demonstrate experimentally that the two-particle coincidence statistics of two bosons can instead be seamlessly tuned from the expected case of suppression to substantial enhancement by an appropriate choice of the observation basis. To this end, our photonic setting leverages birefringent polarization couplers[6] to selectively introduce dissipation in the photons' polarization degree of freedom. Notably, in contrast to[7–9], the mechanism underpinning this this highly unusual behaviour does not act on the individual phases accumulated by pairs of particles along specific paths, but instead allows them to jointly evade losses, while indistinguishable photons are prevented from being simultaneously detected in orthogonal modes. Our findings reveal a new approach to harnessing non-Hermitian settings for the manipulation of multi-particle quantum states and as functional elements in quantum simulation and information processing.**

Energy exchange with the environment is an inescapable feature for any physical system. As such, the fundamental assumption of Hermiticity, as convenient as it is e.g. in the theoretical description of quantum systems, necessarily an approximation only: Non-Hermitian characteristics often tended to be neglected as a matter of course, while the utility of its two sides of gain and loss historically was primarily seen in their capacity for mutual cancellation. Yet, sparked by the ground-breaking work of Bender and Boettcher on parity-time (PT) symmetry[10] and its subsequent adaptation to optical settings[11], a slew of works[12] revealed fascinating features arising from the complex interplay of gain and loss, e.g. non-orthogonal eigenmodes[13], peculiar transport properties[14], loss-induced transparency[15] and unidirectional invisibility[16] as well as exceptional points[16,17] with enhanced response[18], PT-symmetric lasers[19,20] and even light-funnelling[21]. However, while gain and loss are readily incorporated into classical optics as complex-valued refractive index profile, the consequences of changing the number of particles pose fundamental constraints on how these concepts can be brought to bear on genuine quantum states of light[22]. Fortunately, inadvertently introducing additional quantum noise can be avoided by entirely passive configurations. instantiate the desired dynamics. For instance, quasi-parity-time symmetric loss distributions in directional couplers have been shown to systematically accelerate the onset of fully destructive two-photon quantum interference[23]. Along similar lines, lossy beam splitters[7], opaque scattering media[8] or photonically implemented quantum decay processes[9] may enhance two-photon coincidence rates for indistinguishable photon pairs, yielding a peak instead of the conventional dip the HOM experiment, which in Hermitian systems would be indicative of fermionic anti-bunching behaviour.

Here, we experimentally demonstrate that even if loss-induced anti-bunching in the form of a HOM peak is observed, the photons' bosonic characteristics of destructive quantum interference can nevertheless be retained by an appropriate choice of the observation basis. To this end, we implement non-Hermitian two-port couplers based on femtosecond laser-written birefringent waveguides[24] where polarization oscillations of photons play the role of continuous coupling between orthogonal polarization bases,[6] whereas extended ancillary arrays provide the desired amounts of loss[25,26] through polarization-sensitive coupling[27]. Notably, indistinguishable photons are known for being able to partially evade lossy domains in unison,[28] which affects the transition probability of photon pairs depending on their degree of indistinguishability, resulting in an apparent enhancement of the contribution of indistinguishable photon pairs (Fig. 1a). Regardless of the number of transmitted photon pairs however, bunching can be enforced for indistinguishable photons through the choice of the observation basis. As outlined in Fig.

1b, bunched photon pairs cannot be detected via coincidence measurement in specific bases, allowing for a suppression of their registered coincidences $C_{\text{ind}}$ compared to the baseline $C_{\text{dis}}$ of distinguishable photon pairs that remain unaffected (Fig. 1c). We harness a combination of these two effects to induce a continuous transition from enhancement in one polarization basis set to full suppression in its counterpart that has been rotated by 45°. To gain more detailed insights into these dynamics, we record representative HOM patterns for basis orientations associate with visibilities of $v = C_{\text{ind}}/C_{\text{dis}} - 1 < 0$ (i.e. the conventional HOM dip) to peaks ($v > 0$). Furthermore, by experimentally studying the influence of the photon pairs paths through the sample, we show that HOM-interference is actually unaffected by the specific phases accumulated by their constituent particles, revealing an entirely new approach to systematically tailor photon coincidence statistics.

Our setting is based on birefringent waveguides, which can be described with a pair of propagation constants, $\beta_{\text{H}}$ and $\beta_{\text{V}}$ associated with photons polarized horizontally and vertically, in line with the structure's principal axes (see Fig. 2a). As demonstrated recently, such waveguides can be employed as directional coupler in polarization space[6]. Viewed through this lens, they promote a periodic, coherent transfer of excitation between the diagonal state $|D\rangle = (|V\rangle + |H\rangle)/\sqrt{2}$ and the anti-diagonal one $|A\rangle = (|V\rangle - |H\rangle)/\sqrt{2}$ upon propagation along the longitudinal coordinate $z$. While both $|D\rangle$ and $|A\rangle$ evolve with the mean propagation constant $\bar{\beta} = (\beta_{\text{V}} + \beta_{\text{H}})/2$, the rate of coupling between them is governed by the mismatch $\Delta\beta = (\beta_{\text{V}} - \beta_{\text{H}})/2$, i.e. the strength of birefringence. In order to impose dissipation to the system, we employ (spatial) coupling to ancillary waveguide arrays placed in its vicinity (Fig. 2b). While no actual absorption is introduced, the presence of these extended arrangements of likewise birefringent waveguides provide a flexible means to drain light from the states $|H\rangle$ and $|V\rangle$ with polarization-dependent coupling strengths[27] $C_{\text{H}}$ and $C_{\text{V}}$: Once photons have tunnelled out of the target waveguide, the comparatively stronger coupling within the arrays immediately conveys the photons to the sides by means of ballistic discrete diffraction[26]. While the resulting polarization-sensitive loss rates $\gamma_{\text{H}}$ and $\gamma_{\text{V}}$ indeed selectively affect the evolution of the initial amplitudes $\alpha_{\text{H/V}}$ ($\alpha_{\text{H}}^2 + \alpha_{\text{V}}^2 = 1$) of a single-photon state

$$|\psi(z)\rangle = \alpha_{\text{H}} \cdot e^{-\gamma_{\text{H}} z} \cdot e^{-i\beta_{\text{H}} z}|H\rangle + \alpha_{\text{V}} \cdot e^{-\gamma_{\text{V}} z} \cdot e^{-i\beta_{\text{V}} z}|V\rangle \qquad (1)$$

when described as superposition of the polarization eigenstates $|H\rangle$ and $|V\rangle$, the mean loss rate $\bar{\gamma} = (\gamma_{\text{H}} + \gamma_{\text{V}})/2$ represents the balanced global loss rate imposed on the states $|D\rangle$ and $|A\rangle$. In turn, the difference in the individual loss rates $\Delta\gamma = (\gamma_H - \gamma_V)/2$ appears imaginary part in their coupling rate (Fig. 2c).

Horizontally and vertically polarized excitations of the target waveguide indeed are attenuated at markedly different rates, as confirmed by tight-binding simulations (Fig. 3a). Notably, the Zeno dynamics[25] in tight-binding lattices do not yield a purely exponential decay of quantum states, and the resulting deviations at short propagation distances[29] can be accounted for by defining effective loss rates $\gamma_{H/V}$ and their respective mean $\bar{\gamma}$ and mismatch $\Delta\gamma$. Figure 3b shows the observed classical intensity distributions excitations after a propagation length of $z = 15$ cm, where a fraction of 48.3% of vertically polarized light remained in the target waveguide (corresponding to a loss rate of $\gamma_V = 0.02433$ cm$^{-1}$), while the substantially larger coupling in the horizontal polarization component was able to reduce the respective transmission to 4.5% ($\gamma_H = 0.1035$ cm$^{-1}$).

To study the impact on HOM dynamics, a polarization-duplexed two-photon state with one photon each in the diagonal and anti-diagonal states was synthesized from a spontaneous parametric down-conversion (SPDC) photon pair source and injected into the target waveguide. At the other end of the sample, a polarization beam splitter (PBS) served to separate the photons remaining in the target waveguide according to their polarization, and route them to dedicated single-photon counting modules (SPCMs) to register coincidences between the two output ports. Crucially, a half-wave plate (HWP) placed in front of the PBS allowed us to freely choose the orientation $\theta$ of the observation polarization between the horizontal/vertical (H/V, $\theta = 0°$) and diagonal/anti-diagonal (D/A, 45°) cases by an appropriate orientation $\theta/2$ of its fast axis (Fig. 4a). In turn, the characteristic HOM patterns were recorded by varying the two photons' delay $\tau$ at the injection facet between $\tau = 0$ for the indistinguishable configuration and $|\tau| > 1$ ps for the fully distinguishable case.

In a first set of experiments, we set the birefringence of the target waveguide such that the cumulative phase difference along the sample yielded $\Delta\beta \cdot z = 0$ modulo $\pi$, meaning that any observed change to the photons' polarization state at the output is directly associated with the respective loss rates $\gamma_{H/V}$ instantiated by the ancillary arrays. Notably, this configuration removes any impact of the polarization coupling dynamics in the target waveguide on the HOM interference, allowing us to directly characterize the influence of observation basis choice. As shown in Fig. 4b for six representative orientation angles $\theta$ of the half wave plate, the obtained HOM pattern can be continually tuned between the conventional HOM dip regime in the H/V basis ($\theta = 0°$) with a visibility of $v_{HV} = -(94.6 \pm 1.0)\%$, via the case HOM suppression ($v = (2.9 \pm 0.9)\%$ at $\theta = 18°$), to a pronounced HOM peak displaying a visibility of $v_\text{AD} = +(54.1 \pm 0.9)\%$ in the A/D basis ($\theta = 45°$). Note that an ideal HOM peak visibility of $v_\text{AD} = 1$ would be reached in the limiting case of zero transmission in one of the polarizations H or V (see Supplementary Fig. XX for an equivalent realization of this configuration with a high-contrast polarizer).

To gain a more detailed understanding of the interplay between the choice of basis and the polarization-coupling dynamics, we implemented three additional settings: A system with identical losses ($\bar{\gamma} = 0.0639 \text{ cm}^{-1}$, $\Delta\gamma = 0.0396 \text{ cm}^{-1}$) but instead a non-trivial cumulative phase of $\Delta\beta \cdot z \equiv \pi/4$ modulo $\pi/2$, representing a balanced A/D polarization coupler. Furthermore, the "lossless" counterparts ($\bar{\gamma} = 0$, i.e. in the absence of ancillary arrays) for the cases of $\Delta\beta \cdot z = 0$ and $\Delta\beta \cdot z = \pi/4$ were realized. Figure 4c summarizes the dependence of the visibility on basis orientation and polarization coupling (see Supplementary Fig. X for the individual basis-dependent HOM traces). While the A/D basis clearly allows for both HOM dips and HOM peaks to be observed depending on the specific combination of cumulative phase and basis orientation, the H/V basis universally enforces a suppression of the two-photon coincidences ($v_\text{HV} \approx -1$). This prevalence of the conventional HOM dip is inextricably linked to the structure of the two-photon state

$$\Psi_\text{ind}(z) = e^{-2i\bar{\beta}z} \left( \frac{e^{-2i\Delta\beta z} e^{-2\gamma_V z}}{\sqrt{2}} |VV\rangle - \frac{e^{-2i\Delta\beta z} e^{-2\gamma_H z}}{\sqrt{2}} |HH\rangle \right). \quad (2)$$

As it evolves from the injected state $\Psi(0) = (|AD\rangle + |DA\rangle)/\sqrt{2} = (|VV\rangle - |HH\rangle)/\sqrt{2}$ by propagating though the system, the bunching of indistinguishable photons in the H and V polarizations is retained regardless of the propagation phasors $e^{-2i(\bar{\beta} \pm \Delta\beta)z}$ or the specific values of the loss rates $\gamma_{H/V}$. It follows that the indistinguishable photons never coincide in the H/V basis (as indicated in Fig. 1b). In contrast, the two-photon wave function of indistinguishable particles injected as $|A, D\rangle$ evolves as

$$\Psi_\text{dis}(z) = e^{-2i\bar{\beta}z} \left( \frac{e^{-2i\Delta\beta z} e^{-2\gamma_V z}}{2} |V,V\rangle + \frac{e^{-(\gamma_V + \gamma_H)z}}{2} (|V,H\rangle - |H,V\rangle) - \frac{e^{2i\Delta\beta z} e^{-2\gamma_H z}}{2} |H,H\rangle \right) \quad (3)$$

and therefore yields detectable photon pairs in the states $|V, H\rangle$ and $|H, V\rangle$ for any finite $z$. In this vein, the impact of non-Hermiticity on the HOM trace is most apparent in the absence of interference terms ($\Delta\beta \cdot z = 0$ modulo $\pi$). If, in addition, the losses are polarization-independent, i.e. for a trivial loss mismatch ($\Delta\gamma = 0$), the coincidence rates of distinguishable and indistinguishable particles necessarily coincide, and the HOM visibility is suppressed ($v_\text{AD} = 0$). In contrast, in the limiting case of $\gamma_H \to \infty$, only the first terms of Eqs. 2 and 3 remain: Both photons are then V-polarized and register as pairs in the

AD-basis with a probability of $1/4$. Comparing the transmitted amplitudes of the $|VV\rangle$ and $|V,V\rangle$ terms respectively, one notices that they differ by a factor of $\sqrt{2}$: Compared to the baseline of the fully distinguishable case, twice as many indistinguishable photon pairs can in principle be observed, corresponding to a HOM peak with ideal maximal visibility of $v_{\text{AD}} = 1$, whereas finite loss contrasts yield maximal visibilities in the range of $0 < v_{\text{AD}} < 1$ (cf. Fig. 4b, where a maximal visibility of $v_{\text{AD}} = (54.1 \pm 0.9)\%$ was observed for a loss contrast of $\Delta\gamma = 0.0396\ \text{cm}^{-1}$).

As shown in Fig. 4c, the loss contrast is the crucial variable in determining the specific impact that the HOM interference within the sample has on $v_{\text{AD}}$: The polar plot representation of the visibility landscape depending on $\theta$ and $\Delta\beta \cdot z$ illustrate that in the case of vanishing loss contrast ($\Delta\gamma = 0$, left box) only allows for deviations ($v \approx 0$) from strong suppression ($v \approx -1$) in a narrow region surrounding the AD-basis ($\theta = 45°$) for trivial cumulative phases ($\Delta\beta \cdot z \equiv 0$). In contrast, a non-trivial loss mismatch ($\Delta\gamma = 0.0396\ \text{cm}^{-1}$, right box) leads to nearly $\Delta\beta \cdot z$-independent visibilities for any given basis orientations $\theta$. This calculated behaviour is confirmed by the observed visibilities. These measurements clearly show that the requirement of $\Delta\beta \cdot z \approx \pi/4$ for quantum interference in the lossless case can be overcome by introducing non-trivial loss contrasts. This suggests that a new mechanism for tailoring two-photon coincidences has indeed been identified.

In conclusion, our findings outline a new approach to shape two-photon coincidence statistics through the interplay of non-Hermitian systems and multiple-particle quantum mechanics. To this end, the polarization degree of freedom allows both for a seamless adjustment of the observation basis and for the implementation of complex loss profiles. Notably, the techniques described here can be readily combined with the degrees of freedom afforded by the three-dimensional spatial arrangement of birefringent waveguides[24], specific exchange symmetries of the input states[30] and waveguide structures in higher synthetic dimensions[6], and open up a number of fascinating opportunities: Probabilistic quantum gates stand to benefit greatly from the capability to introduce sophisticated loss realization in integrated optical platforms. Arbitrary quantum-optical transformations such as the singular value decomposition[31] can be emulated in photonic model systems, and even experiments on the evolution of multiple-particle quantum states in complex non-Hermitian systems become experimentally accessible.

**Methods and Materials**

*Waveguide fabrication and birefringence characterization*

We employ the femtosecond laser direct writing technique[24] to inscribe systems of evanescently coupled single-mode waveguides. To this end, ultrashort laser pulses with $270\,\text{fs}$ duration at a carrier wavelength of $\lambda = 517\,\text{nm}$ and a repetition rate of $333\,\text{kHz}$ from a fibre amplifier system (Coherent Monaco) were focused into the bulk of $150\,\text{mm}$ long fused silica samples (Corning 7980) through a $50\times$ objective ($\text{NA} = 0.60$), where upon waveguides are formed along desired trajectories by translating the sample with a precision positioning system (Aerotech ALS180).

The combination of elliptical material modifications with residual stress fields from the rapid quenching immediately following the inscription pass imbues the laser-written waveguides with an inherent birefringence, the strength of which can be tuned via the writing speed[32]. We characterized the waveguides using classical light and crossed polarizers[33]: Having placed the sample between two polarizing beam splitters and simultaneously rotating them by an angle $\varphi$ (while maintaining their respective crossed orientation), the ratio of the transmitted ($I_\text{T}$) and total intensities ($I_\text{total}$) was measured with two photo diodes and evaluated according to the theoretical model

$$\frac{I_\text{T}}{I_\text{total}} = \sin^2(2\varphi) \cdot \sin^2(\Delta\beta \cdot z),$$

allowing for the relevant values of $0$ and $\pi/4$ of the cumulative phase difference $\Delta\beta \cdot z$ to be identified by $\max(I_\text{T}/I_\text{total}) = 0.5$ and $\max(I_\text{T}/I_\text{total}) = 0$ up to a multiples of $\pi/2$ and $\pi$, respectively.

For our experiments, we chose a separation of $27.5\,\mu\text{m}$ between the target waveguide and the ancillary arrays, corresponding to coupling strengths of $C_\text{H} = 0.154\,\text{cm}^{-1}$ and $C_\text{V} = 0.065\,\text{cm}^{-1}$ of the polarization eigenstates to their respective "sinks". In order to ensure that photons, once extracted from the target guide, are swiftly transported away, an array pitch of $20\,\mu\text{m}$ was used to set the intra-array coupling strengths to $0.551\,\text{cm}^{-1}$ and $0.335\,\text{cm}^{-1}$ for the horizontal and vertical polarizations, respectively.

*Bi-photon creation and detection*

For the generation of photon pairs, we used a type I SPDC source (see Fig. 4a). A BiBO crystal is pumped by a $100\,\text{mW}$ continuous-wave laser diode (Coherent OBIS) at $\lambda = 407\,\text{nm}$, producing wavelength-degenerate horizontally polarized photon pairs at $\lambda = 814\,\text{nm}$, which are subsequently collected by polarizing fibres and routed to the sample via a polarization combiner with adjustable principal axes of the output fibre. A variable delay $\tau$ between the arrival times of the two photons at the sample is implemented by means of a motorized translation stage (PI). The degree of indistinguishability of the photons was characterized by recording the HOM dip with a fibre-based beam splitter, yielding a visibility of up to $v = -(97.2 \pm 1.1)\%$. In order to record coincidences, the single photons at the output of the PBS are detected with Single Photon-Counting Modules (Excelitas Technologies) whose signals are processed with a correlation card.

**Acknowledgements:**

The authors would like to thank C. Otto for preparing the high-quality fused silica samples used in this work.

**Author contributions:**

ME fabricated the samples and carried out the measurements. All authors jointly interpreted the measured data and co-wrote the manuscript.

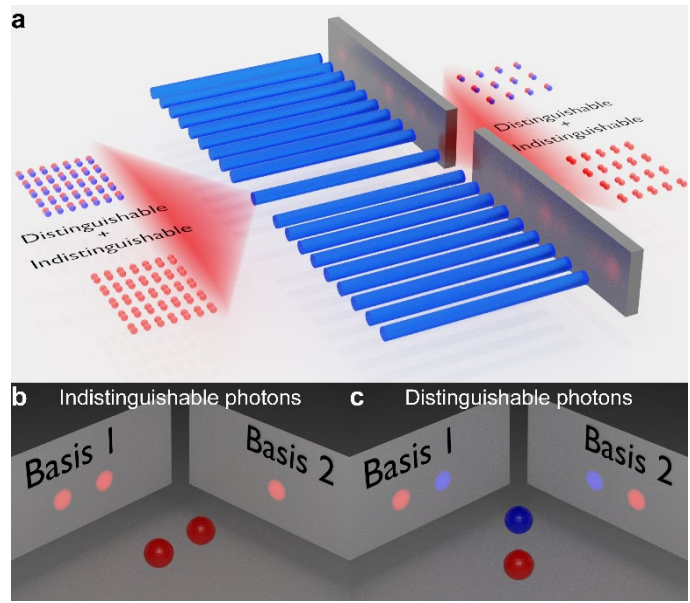

**Figure 1: Photon pairs' propagation and detection in lossy systems. A.** The same quantity of distinguishable (red and blue) and indistinguishable (both in red) photon pairs are launched in a single waveguide. Lossy regions, represented by the coupling to the surrounding waveguides, lead to a stronger decrease in the distinguishable photon pairs' transition probability so that indistinguishable photon pairs appear enhanced. **b,** Illustration of observation-dependent coincidence detection for indistinguishable photon pairs. For an observation in basis 1, the photons are detected in different states and which makes it possible to measure them as coincidences. In basis 2, the photons appear in the same state (bunched) and cannot be detected via coincidence measurement between different states. **c,** Distinguishable photons do not appear in bunched states and can be measured as coincidences independent on the measurement basis.

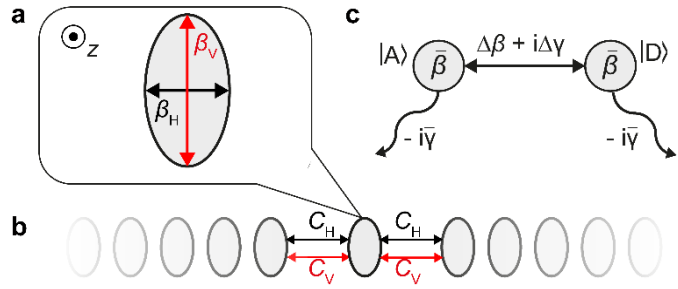

**Figure 2: Loss implementation in a polarization coupler. a,** A birefringent waveguide with the propagation constants $β_H$ and $β_V$ on his horizontally and vertically orientated principal axes is **b,** coupled to two ancillary arrays consisting of tightly coupled waveguides. Polarization-sensitive coupling strengths $C_H$ and $C_V$ between the target to the first sink waveguide as well as different couplings between the sink waveguides enable polarization-dependent loss rates $γ_H$ and $γ_V$ for the target (central) waveguide. **c,** Illustration of the single-particle Hamiltonian in the AD-polarization basis. $\bar{β} = (β_V + β_H)/2$ and $Δβ = (β_V - β_H)/2$ directly result from the waveguide's birefringence, whereas the imaginary part of the coupling strength $Δγ = (γ_H - γ_V)/2$ and globas loss $\bar{γ} = (γ_H + γ_V)/2$ arise from the loss rates' mismatch.

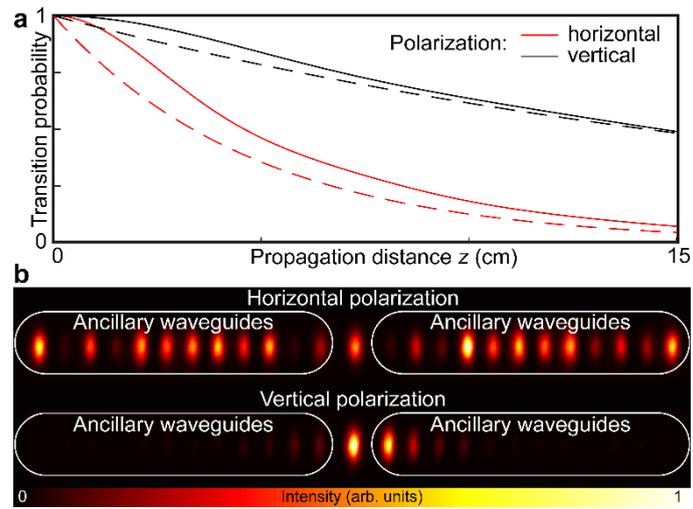

**Figure 3: Light propagation in the sample. a,** Transition probabilities for horizontal (red) and vertical (black) polarization excitation in the target waveguide resulting from the a tight-binding simulation of the waveguide structure with the given coupling strengths. The dashed lines represent the transition probability for the effective decay rates. **b,** Classical intensity distribution of the output after $z = 15$ cm observed with a CCD camera when the target waveguide is illuminated at $\lambda = 814\ nm$. For an excitation in the horizontal polarization, only $4.5\%$ intensity remain in the target waveguide, compared to $48.3\%$ for the vertical polarization. The corresponding loss rates are $\gamma_\text{H} = 0.1035\ \text{cm}^{-1}$ and $\gamma_\text{V} = 0.02433\ \text{cm}^{-1}$.

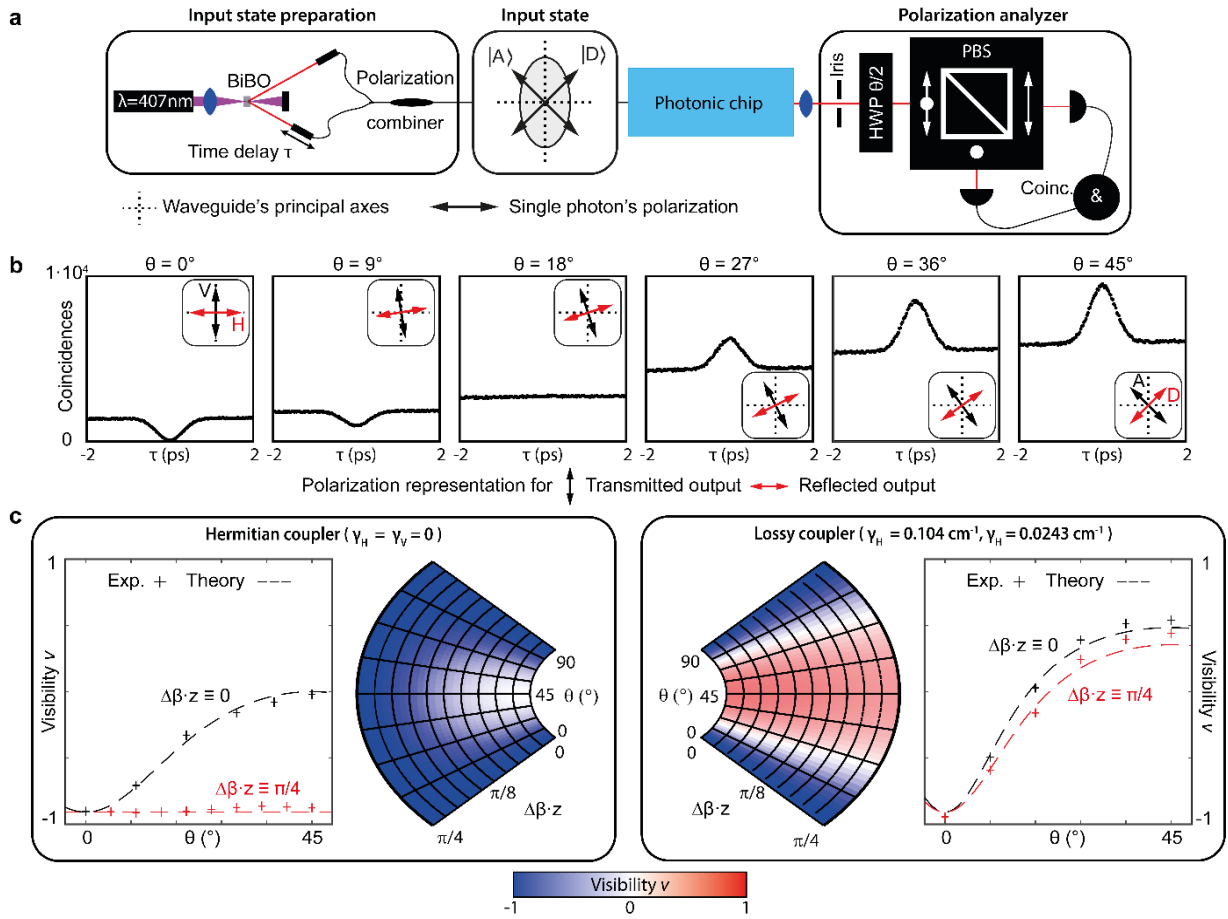

**Figure 4: Switching between HOM dip and peak. a,** Experimental setup for the HOM experiment. The photon pairs from a type-I-SPDC source are collected with polarizing fibres. In turn, a polarization combiner serves to prepare an input state with one photon polarized diagonally and antidiagonally with respect to the waveguide's principal axes. After a subsequent propagation in the sample, the remaining photons in the target waveguide are analysed with a polarization beam splitter (PBS). The photons detected with single photon-counting modules at the PBS' transmitting and reflecting port, record coincidences in the orthogonal polarization bases adjusted via the half-wave plate's (HWP) principal axes orientation $\theta/2$ towards the vertical axis. **b,** HOM patterns recorded in different polarization basis that are subsequently rotated from the HV-basis ($\theta = 0°$) to the AD-basis ($\theta = 45°$). The basis representation for the transmitted and reflected output ports of the PBS are shown with black and red arrows in the HV frame of reference (dashed lines) **c,** Predicted (dashed lines) and recorded (plusses) HOM visibilities according to the observation basis orientation $\theta$ for the Hermitian (left) and lossy (right) coupler with the accumulated phases $\Delta\beta \cdot z = 0$ modulo $\pi$ (black) and $\Delta\beta \cdot z = \pi/4$ modulo $\pi/2$ (red). Colour-encoded HOM visibilities are shown in the corresponding polar plots (centre) in dependence of basis orientation $\theta$ and accumulated phase $\Delta\beta \cdot z$.